\documentclass[aps,twocolumn,showpacs,preprintnumbers,nofootinbib,superscriptaddress]{revtex4}
\usepackage{amsmath} \usepackage{graphicx} \usepackage{amsfonts}
\usepackage{array} \usepackage{amsthm} \usepackage{bm} 

\usepackage[breaklinks]{hyperref}
\usepackage{color}

\newcommand{\nn}{\nonumber}
\newcommand{\be}{\begin{equation}}
\newcommand{\ee}{\end{equation}}
\newcommand{\ba}{\begin{eqnarray}}
\newcommand{\ea}{\end{eqnarray}}
\newcommand{\bal}{\begin{align}}
\newcommand{\eal}{\end{align}}

\newcommand{\e}{{\rm e}}
\newcommand{\dd}{{\rm d}}

\newcommand{\bb}{\bibitem}

\newcommand{\al}{\alpha}

\newcommand{\bt}{\beta}

\newcommand{\ro}{\rho}
\newcommand{\ep}{\epsilon}
\newcommand{\si}{\sigma}
\newcommand{\ta}{\theta}
\newcommand{\Ta}{\Theta}

\newcommand{\De}{\Delta}

\newcommand{\de}{\delta}

\newcommand{\bw}{\begin{widetext}}
\newcommand{\ew}{\end{widetext}}

\def\abh{black hole }

\def\aBH{black hole}

\def\Sd{Schwarzschild }

\begin{document}
\title{Confined-exotic-matter wormholes with no gluing effects -- Imaging supermassive wormholes and black holes}

\author{Mustapha Azreg-A\"{\i}nou}
\affiliation{Ba\c{s}kent University, Faculty of Engineering, Ba\u{g}l\i ca Campus, 06810 Ankara, Turkey}


\begin{abstract}
We classify wormholes endowed with redshift effects and finite mass into three types. Type I wormholes have their radial pressure dying out faster, as one moves away from the throat, than any other component of the stress-energy and thus violate the least the local energy conditions. In type II (resp. III) wormholes the radial and transverse pressures are asymptotically proportional and die out faster (resp. slower) than the energy density. We introduce a novel and generalizable method for deriving, with no cutoff in the stress-energy or gluing, a class of each of the three wormhole types. We focus on type I wormholes and construct different asymptotically flat solutions with finite, upper- and lower-bounded, mass $M$. It is observed that the radial pressure is negative, and the null energy condition is violated, only inside a narrow layer, adjacent to the throat, of relative spacial extent $\epsilon$. Reducing the relative size of the layer, without harming the condition of traversability, yields an inverse square law of $\epsilon$ versus $M$ for supermassive wormholes. We show that the diameter of the shadow of this type I supermassive wormhole overlaps with that of the black hole candidate at the center of the Milky Way and that the recent derivation, using the up-to-date millimeter-wavelength very long baseline interferometry made in Astrophys. J. \textbf{795} 134 (2014) [arXiv:1409.4690], remains inconclusive.

We show that redshift-free wormholes, with positive energy density, have one of their barotropic equations of state in the phantom regime (at least in the region adjacent to the throat), have their stress energy tensor traceless, and are anisotropic. They are all type III wormholes having their variable equations of state approaching 1 and $-1$ at spatial infinity. We also introduce a new approach for deriving new redshift-free wormholes.
\end{abstract}


\maketitle

\section{Introduction\label{secI}}

How exotic is exotic matter? Do young and old galaxies harbor exotic matter? So far there has been no simple or advanced theory about exotic matter nor a prediction and all we know about it is its mathematical definition: it violates our perception of energy. That is, if an observer measures some negative local amount of energy density, we say that that corresponds to exotic matter.

The other thing we know about exotic matter is its possible wormhole sustainability~\cite{MT,V}.  While they are of exotic nature, wormholes may interact with ordinary matter and may be indirectly observed through the effects they have on light and particle paths as well as on fields~\cite{Chet,Az,gf,sh,gf2}, on falling hot objects and spots~\cite{hs} and so on.

The field equations of classical general relativity do not fix the topology of its solutions nor do they fix the amount of exotic matter needed to sustain the throat of a wormhole. Quantum effects allow for violations of the local and averaged~\cite{anec} null energy condition (NEC) and might be used to support and stabilize wormholes.

Since exotic matter remains still a mystery, workers, using different techniques, have ever strived hard to derive wormhole solutions that minimize its use~\cite{MT,V} and~\cite{m1}-\cite{m8}. To the best of our knowledge, no classification of wormholes has been performed so far. Observers of events are usually located far away from the sources, say, at spatial infinity, where observable entities may behave differently. The only distinctions among wormholes, which are widely used by workers, are finiteness of the mass, traversability, and stability. Other observable entities that may distinguish between wormholes are the components of the stress energy tensor (SET). It is the duty of this paper to perform this classification based on the relative behavior of the components of the SET at spatial infinity.

Another, but implicit, classification of wormholes concerns redshift-free wormholes and wormholes endowed with it. The radial and transverse pressures of redshift-free wormholes, with positive energy density and finite mass, behave the same way at spatial infinity, so there is no classification added for these solutions. This fact could be announced as a uniqueness theorem. This is no longer the case for wormholes endowed with redshift effects where three types of solutions, having finite mass, emerge.

The classification of wormholes motivates a new mathematical quest for theoretical wormholes fueled by the recent activities~\cite{sh,hs} to whether the observations of the shadow or hot spots are able to distinguish between a supermassive black hole (SMBH), located at Sagittarius A$^\star$ (Sgr A$^\star$), and a supermassive wormhole (SMWH). Questioning if the SMBH candidate at the center of the Milky Way is a SMWH is right but trying to answer it is hard. In fact, we have noticed that the wormhole solutions used in these investigations are the types that demand the most exotic matter. We will show that it is possible, without using the cut and paste technique, to construct their counterparts which violate the least the NEC and yield a value $46\,\mu\text{as} \text{ --- } 54 \,\mu\text{as}$ for the diameter of the shadow.

This is the same value derived very recently~\cite{Sgr}, using the millimeter-wavelength very long baseline interferometry (VLBI). The image of the emission surrounding the SMBH candidate in the center of the Milky Way reveals, at 1.3 mm VLBI, the same features of general relativity including that of a shadow of diameter $\sim 50\,\mu\text{as}$. Knowing that Sgr A$^\star$ along with M87 are on the list of the main targets of the Event Horizon Telescope~\cite{EHT}, the sensitivity of these measurements will increase with the inclusion of Atacama Large Millimeter/Submillimeter Array VLBI-station~\cite{VLBI}.

In Sec.~\ref{secfe} we review the field equations ant the local energy condition's (LEC's). We specialize to solutions having finite mass and positive energy density and derive some general formulas. In Sec.~\ref{secs} we consider redshift-free static wormholes and construct by a new procedure some new exact solutions. Those redshift-free wormholes, with positive energy density, have one of their barotropic equations of state in the phantom regime (at least in the region adjacent to the throat), have their stress energy tensor traceless, and are anisotropic. In Sec.~\ref{secr} we focus more on solutions with variable redshift function and constant finite mass. We classify them into three types I, II, and III. Type I (respectively type III) wormholes violate the least (respectively the most) the LEC's. The importance of type I and type III solutions is that they can be used by distant observers for testing hypotheses and in computer simulations. We introduce a 3-parameter approach to derive, without gluing, a class of each of the three wormhole types. The approach splits into two directions, in the one of which only one parameter remains free, and in the other one two parameters remain free to confine the exotic matter. We discuss the violations of the LEC's and traversability.

Sec.~\ref{secsd} is devoted to an application. First, we show that the wormhole solution that has been used~\cite{sh} for evaluating the shadow of the SMBH candidate is type III. We use, instead, a type I solution and show that the evaluation of the shadow is inconclusive. Said otherwise, the outcome of the observation is such that (a) the candidate might either be a (\Sd or Kerr) SMBH or a type I SMWH, (b) the candidate is a type III SMWH with relatively large amounts of exotic matter in the center of the galaxy. Based on the recent results of Ref~\cite{Sgr}, this last possibility is ruled out.

In Sec.~\ref{secg} we show how the approach introduced in Sec.~\ref{secr} can be generalized and provide two more wormhole solutions. We conclude in Sec.~\ref{secc}.

\section{Field equations and LEC's \label{secfe}}

The metric of a static, spherically symmetric, wormhole is better brought to the form~\cite{MT}
\begin{equation}\label{b1}
    \dd s^2=A(r)\dd t^2-\frac{\dd r^2}{1-b(r)/r}-r^2\dd \Omega^2,
\end{equation}
in \Sd coordinates. The throat is located at $r=r_0>0$ and it corresponds to the minimum value of $r^2$. We assume symmetry of the two asymptotically flat regions, which particularly implies that if the mass of the wormhole is finite then it is the same as seen from both spatial infinities. The metric~\eqref{b1}, representing a wormhole solution ($A>0$ for $r\geq r_0$), is exempt from any singularity, particularly the curvature $\mathcal{R}$ and Kretschmann $R_{\al\bt\mu\nu}R^{\al\bt\mu\nu}$ scalar invariants are regular everywhere on the throat and off it
\begin{equation}\label{inv}
\mathcal{R}=\frac{P_{C}}{2r^2A^2},\quad  R_{\al\bt\mu\nu}R^{\al\bt\mu\nu}=\frac{P_{K}}{4r^6A^4},
\end{equation}
where ($P_C,P_K$) are polynomials in $A(r)$ and its first and second derivatives, $b(r)$ and its first derivative, and $r$.

Besides the constraint $A>0$ for $r\geq r_0$, the functions $A$ and $b$ are further constrained by~\cite{MT,V}
\begin{align}
&\lim_{r\to\infty}A=\text{finite}=1,\nn\\
&b<r\text{ if }r>r_0\;\text{ and }\;b(r_0)=r_0,\nn\\
\label{b2}&\lim_{r\to\infty}(b/r)=0,\\
&rb'<b\;(\text{in the region adjacent to the throat}),\nn\\
&b'(r_0)\leq 1.\nn
\end{align}
Notice that $rb'=b$ may hold on the throat. The value of the limit in the first line~\eqref{b2} is set to 1 by rescaling $A$ and redefining $t$. For a wormhole solution, the shape function $b$, which is positive on the throat, need not preserve the same sign\footnote{When $b>0$ for all $r\geq r_0$, the two-dimensional sections of the wormhole can be entirely embedded in a three-dimensional Eucledian space since, in this case, the rhs of the embedding Eq.~(27) of Ref.~\cite{MT} is always real.} on the whole range of $r$. In the case where $b$ may have both signs, the fourth line~\eqref{b2} is violated, not in the region adjacent to the throat (where $b>0$), rather in the region(s) where $b'\geq 0$ and $b<0$. This results in a wormhole solution with an effective mass inside the radius $r$ that is negative whenever $b<0$ [Eq.~(11.42) of Ref.~\cite{V}].

The first and third lines in~\eqref{b2} ensure asymptotic flatness. The proper radial distance is defined by
\begin{equation}\label{prd}
\ell\equiv\int_{r_0}^r\sqrt{-g_{rr}(\bar{r})}\dd \bar{r}=\int_{r_0}^r\frac{\dd \bar{r}}{\sqrt{1-b(\bar{r})/\bar{r}}}.
\end{equation}
The first constraint in the second line~\eqref{b2} ensures that $\ell$ is real and the second one ensures that the throat $r_0$ is a minimum value of $r(\ell)$. The remaining constraints, fourth and fifth lines in~\eqref{b2}, ensure that $r(\ell)$ is an increasing, convex (concave up), function of $\ell$; that is, as one moves away from the throat $r(\ell)$ increases and turns upward (here $\ell$ represents a horizontal axis and $r$ a vertical one).

The constraints~\eqref{b2} hold even if the mass of the wormhole is not finite. If the latter is finite, we have the further constraint
\begin{equation}\label{b3}
\lim_{r\to\infty}b\equiv b_{_\infty}=2GM=2M.
\end{equation}

The SET is usually taken anisotropic of the form~\cite{MT,V} $T^{\mu}{}_{\nu}={\rm diag}(\ro(r),-p_r(r),-p_t(r),-p_t(r))$, $\ro$ being the energy density and $p_r$ and $p_t$ are the radial and transverse pressures. The filed equations $G^{t}{}_{t}=8\pi T^{t}{}_{t}$, $G^{r}{}_{r}=8\pi T^{r}{}_{r}$, and the identity $T^{\mu}{}_{r;\mu}\equiv 0$ yield, respectively
\begin{align}\label{b4}
&b'=8\pi r^2\ro ,\nn\\
&(\ln A)'=\frac{8 \pi  r^3p_r +b}{r (r-b)},\\
&4 p_t=4 p_r+2 r p_r{}'+r(p_r+\rho )(\ln A)',\nn
\end{align}
where a prime denotes derivation with respect to $r$.

The SET is subject to the requirements of the LEC's, known as null, weak (WEC), strong (SEC), and dominant (DEC) conditions. These requirements read, respectively~\cite{V}
\begin{align}\label{ec}
&\text{NEC: }& &\ro+p_r\geq0,\,\ro+p_t\geq0, \nn\\
&\text{WEC: }& &\ro\geq0,\,\ro+p_r\geq0,\,\ro+p_t\geq0, \nn\\
&\text{SEC: }& &\ro+p_r\geq0,\,\ro+p_t\geq0,\,\ro+p_r+2p_t\geq0, \nn\\
&\text{DEC: }& &\ro\geq0,\,p_r\in[-\ro,\ro],\,p_t\in[-\ro,\ro] .
\end{align}

One of our main purposes in this work is to construct wormhole solutions that violate the least the LEC's. These are the more realistic wormholes in the framework of the theory of general relativity or its extended theories. From now on and throughout this paper, we specialize to wormholes having a positive energy density $\ro$, this already frees us from concern with one of the constraints of the LEC'S~\eqref{ec}. Working with $\ro\geq 0$ is a common approach followed by workers in this field (see for instance~\cite{MT,LPR,LPR1,LPR2}). If the mass is finite, which is the case in which we will be interested most, and $\ro\geq 0$, Eqs.~\eqref{b2} and~\eqref{b3} along with the first line~\eqref{b4} yield
\begin{equation}\label{cm}
    2M-r_0=8\pi\int_{r_0}^{\infty}r^2\ro\,\dd r \geq 0\qquad (\text{for }\ro\geq 0).
\end{equation}
This implies
\begin{equation}\label{cm1}
 M\geq r_0/2,
\end{equation}
for wormholes with finite mass and positive or null energy density. The saturation is attained only for $\ro = 0$: $M= r_0/2$. This sets a lower limit for the mass of wormholes whose energy density is everywhere positive or null.

Two other conclusions that result from $\ro >0$ are:
\begin{align}
\label{cm1a}&b\geq r_0\qquad (\forall\,r\geq r_0),\\
\label{cm1b}&x\equiv 8\pi r_0{}^2\ro_0\leq 1 \qquad [\ro_0\equiv \ro(r_0)>0].
\end{align}
The positiveness of $b$, when $\ro >0$, results from application of the second line~\eqref{b2} and the first line~\eqref{b4}. The inequality in~\eqref{cm1b} results from application of the fifth line~\eqref{b2} and the first line~\eqref{b4}.

The condition that the mass is finite defines the asymptotic behavior of $\ro$. In order for the integral in~\eqref{cm} to converge, $\ro >0$ must behave as
\begin{equation}\label{cm2}
 \ro \sim  \ro_{_\infty}r^{-3-\si}\;\;\text{ as }\;\; r\to\infty\qquad (\si >0).
\end{equation}

There does not seem to be an upper limit for the mass that is valid for all wormholes having finite mass and positive energy density. As we shall see in the subsequent sections, if such an upper limit exits it will depend on the whole expression of $\ro$; that is, the near throat~\eqref{cm1b} and asymptotic~\eqref{cm2} behaviors are not sufficient to fix an upper limit for the mass.

\section{Redshift-free static wormholes\label{secs}}

If no redshift effects occur ($A=1$) Eqs.~\eqref{b4} take the forms
\begin{align}\label{b6}
&b'=8\pi r^2\ro ,\nn\\
&8 \pi  r^3p_r +b=0,\\
&2 p_t=2 p_r+r p_r{}'.\nn
\end{align}
Since the the energy density and the pressures depend only on $r$, we can always assume two barotropic equations of state of the form
\begin{equation}\label{b5}
p_r(r)=\al(r)\ro(r),\quad p_t(r)=\bt(r)\ro(r).
\end{equation}
These barotropic assumptions are valid for any static, spherically symmetric, solution be it redshift-free or other. Since we are interested in the case of positive $\ro$, the inequality~\eqref{cm1a} applies: $b\geq r_0>0$. The second line~\eqref{b6} shows that $p_r$ and $\al$ are negative for all $r\geq r_0$. Combining the last two lines~\eqref{b6} yields
\begin{equation}\label{b7}
    p_t=\frac{b-rb'}{16\pi r^3},
\end{equation}
which is positive by the fourth line~\eqref{b2} at least in the region adjacent to the throat. Thus, $\bt$ and the dimensionless anisotropy parameter~\cite{MH,V2} $\De\equiv (p_t-p_r)/\ro =\bt-\al$ are both positive (at least in the region adjacent to the throat).

Using~\eqref{b5}, the first two lines~\eqref{b6} and~\eqref{b7} yield
\begin{align}
&b(r)=r_0\exp\Big(-\int_{r_0}^{r}\frac{\dd \tilde{r}}{\tilde{r}\al(\tilde{r})}\Big),\nn\\
\label{b8}&\ro=b'/(8\pi r^2),\\
&2\bt=-(\al+1),\nn
\end{align}
where we have used the second line~\eqref{b2}. We see that in the absence of redshift effects the knowledge of $\al(r)$ suffices to determine all the necessary functions ($b,\ro,p_r,p_t$). This consists the method of resolution we introduce to construct wormhole solutions with no redshift effects. Equivalently, one may use $\bt$, instead of $\al$, to determine all the other functions.

The third line~\eqref{b8} implies that the SET is traceless $\ro+p_r+2p_t=0$.

The coefficient $\bt(r)$ being positive at least in the region adjacent to the throat, the last line~\eqref{b8} results in
\begin{equation}\label{b9}
    \al(r)<-1\;\;(\text{in the region adjacent to the throat}),
\end{equation}
which lies in the phantom regime (at least in the region adjacent to the throat). Notice that in this regime it is not possible to have $\bt=\al$, that is, an isotropic solution, since in this case the third line~\eqref{b8} would imply $\al=-1/3$, which is not allowed by~\eqref{b9}. The conclusion is even stronger than that: For a given solution, there is no sphere of radius $r\geq r_0$ where all the components of the pressure are equal to each other. A similar conclusion concerning gravastars was drawn in~\cite{V2} where it was shown that such objects cannot be perfect fluids.

Using~\eqref{b8} and~\eqref{b9} it is straightforward to show that most of the local (null, weak, strong, and dominant) energy conditions~\cite{V} are violated at least in the region adjacent to the throat, for $\ro>0$ implies $\ro+p_r=(1+\al)\ro<0$ and $p_r \notin [-\ro,+\ro]$; the condition $p_t \in [-\ro,+\ro]$ is satisfied only if $\bt\leq 1$ ($-3\leq\al<-1$). The constraint $\ro+p_t=(1+\bt)\ro>0$ is satisfied at least in the region adjacent to the throat and $\ro+p_r+2p_t=0$ is satisfied everywhere.

There is a variety of factors $\al(r)$ leading to closed-form expressions for all the functions ($b,\ro,p_r,p_t$). These can be easily seen from the first line~\eqref{b8} and are investigated in the following two subsections. So, in the remaining part of this section, we fix the expression of $\al(r)$ and use~\eqref{b8} to determine simple expressions for the functions ($b,\ro,p_r,p_t$) and the metric.

\subsection{$\pmb \al$ is constant}

By~\eqref{b9} we see that the only possibility where $\al$ is constant is the case $\al = \text{constant}<-1$. Let $\nu\equiv -1/\al$ yielding $0<\nu<1$. The metric and the necessary functions read
\begin{align}\label{s1}
&\dd s^2=\dd t^2-\frac{\dd r^2}{1-(r_0/r)^{1-\nu}}-r^2\dd \Omega^2,\nn\\
&b=r_0(r/r_0)^{\nu},\quad \ro=\frac{\nu}{8\pi r_0{}^2(r/r_0)^{3-\nu}},\\
&p_r=-\ro/\nu,\quad p_t=\frac{1-\nu}{2\nu}\ro\qquad (0<\nu<1).\nn
\end{align}
It is easy to check that the limits of ($\ro,p_r,p_t$), as $r\to\infty$, vanish but that of $b$ diverges, so the mass is infinite. It is also easy to check that all the constraints~\eqref{b2} are satisfied. The special case $\nu=1/2$ was discussed in Ref.~\cite{MT}. This solution has been rederived in~\cite{Kalam,LPR}.

\subsection{$\pmb{\al+1\propto -(r/r_0)^{\mu}}$, $\pmb{\mu>0}$}

Another closed-form solution is derived taking $\al+1\propto -(r/r_0)$. To satisfy~\eqref{b9} we may choose, for the sake of simplicity, the constant of proportionality positive everywhere. Thus, we write $\al$ as
\begin{equation}\label{s2}
\al=-1-k^2\frac{r}{r_0}\;\text{ and }\;k^2>0.
\end{equation}
Direct integration yields
\begin{equation*}
    b=\frac{(1+k^2)r_0r}{k^2r+r_0}.
\end{equation*}
Since $b_{_\infty}=(1+k^2)r_0/k^2$ is finite we introduce the mass parameter defined in~\eqref{b3}: $2M=(1+k^2)r_0/k^2$. This implies the general result~\eqref{cm1}: $M>r_0/2$. In terms of ($M,r_0$) we obtain the following solution
\begin{align}\label{s3}
&\dd s^2=\dd t^2-\Big(1-\frac{2 M}{r+2 M-r_0}\Big)^{-1}\dd r^2-r^2\dd \Omega^2,\nn\\
&b=\frac{2 M r}{r+2 M-r_0},\quad \ro=\frac{M (2 M-r_0)}{4 \pi  r^2 (r+2 M-r_0)^2}>0,\\
&p_r=-\frac{M}{4 \pi  r^2 (r+2 M-r_0)},\quad p_t=\frac{M}{8 \pi  r (r+2 M-r_0)^2}.\nn
\end{align}

The components of the SET vanish at spatial infinity. The constraints~\eqref{b2} are all satisfied. For instance, the last two lines~\eqref{b2} read, respectively
\begin{equation*}
-\frac{2Mr^2}{(r+2 M-r_0)^2}<0\;\;(\forall\;r\geq r_0),\quad -\frac{r_0}{2M}<0.
\end{equation*}

The above solution generalizes easily to the case\footnote{Solutions of the form $\al=-1-k^2(r_0/r)^{\mu }$ with $\mu>0$ do not satisfy the third line~\eqref{b2}.} $\al+1\propto -(r/r_0)^{\mu}$ and $\mu>0$
\begin{align}\label{s4}
&\dd s^2=\dd t^2-\Big(1-\frac{2 M}{\mathcal{R}^{1/\mu }}\Big)^{-1}\dd r^2-r^2\dd \Omega^2,\nn\\
&b=\frac{2 M r}{\mathcal{R}^{1/\mu }},\quad
\ro=\frac{M (2^{\mu } M^{\mu }-r_0{}^{\mu })}{4 \pi  r^2
\mathcal{R}^{(\mu +1)/\mu }}>0,\\
&p_r=-\frac{M}{4 \pi  r^2 \mathcal{R}^{1/\mu }},\quad p_t=\frac{M}{8 \pi  r^{2-\mu } \mathcal{R}^{(\mu +1)/\mu }},\nn\\
&\mathcal{R}\equiv r^{\mu }+2^{\mu } M^{\mu }-r_0{}^{\mu },\quad M>r_0/2,\quad \mu>0.\nn
\end{align}

According to the analysis made in~\cite{MT}, these wormholes are traversable. At spatial infinity the energy density dies out as fast as $r^{-3-\mu}$ and the pressures as $r^{-3}$ where $\mu$ is an arbitrary positive constant. This behavior is general and applies to all redshift-free static wormholes with finite mass parameter. In fact, if $\ro\sim \ro_{_\infty}r^{-3-\mu}$ ($\mu>0$) as $r\to\infty$, then the first, second, and third lines in~\eqref{b6} yield, respectively, $b\sim b_{_\infty}-8\pi\ro_{_\infty}r^{-\mu}/\mu$, $p_r\sim -b_{_\infty} r^{-3}/(8\pi)$, and $p_t\sim b_{_\infty} r^{-3}/(16\pi)$. The solutions~\eqref{s4}, as well as the special case~\eqref{s5}, are the simplest ones with these properties. Since the violations of the LEC's are attributable to $p_r$, which is negative, this dashes any hope for obtaining redshift-free solutions with $p_r$ dying out faster than $r^{-3}$.

Now, we consider the limiting case $M=r_0/2$. We obtain the wormhole solution~\cite{MT}
\begin{align}\label{s5}
&\dd s^2=\dd t^2-\Big(1-\frac{2 M}{r}\Big)^{-1}\dd r^2-r^2\dd \Omega^2,\nn\\
&b=2M,\quad M=r_0/2,\quad\ro\equiv 0,\\
&p_r=-\frac{M}{4 \pi  r^3},\quad p_t=\frac{M}{8 \pi  r^3},\nn
\end{align}
which can also be derived from~\eqref{b6} and~\eqref{b7} taking $b=\text{ constant }=2M$ [It is also derived from~\eqref{s4} taking the limit $\mu\to\infty$]. With $\rho\equiv 0$ and $p_r<0$, this represents the most exotic matter distribution. This is not a \Sd wormhole since the latter has $\ro=p_r=p_t\equiv 0$.

Had we assumed the fluid isotropic, such a solution would not exist even in the full regime where $A$ is not constant. In fact, a solution which behaves at spatial infinity as $\ro\sim r^{-3-\si}$ and $p_r=p_t\sim r^{-3}$ yields a nonasymptotically flat solution.

The solutions derived in this section have their pedagogical values and will be added to the long list of solutions used for pedagogical purposes (see for instance~\cite{MT,LPR,LPR1,LPR2}, \cite{az3}-\cite{refw}). Except the solution~\eqref{s1}, which has been derived elsewhere, the solution~\eqref{s3} and its generalization~\eqref{s4} are new and have simple structures. Because of these latter properties, some of these pedagogical solutions may find their way to applications: they may have a potential use in computer simulations and/or in testing hypotheses. This was the case with the rotating wormhole derived for pedagogical purposes in Ref.~\cite{teo}, which was used in Ref.~\cite{col} to investigate the high energy collision of two particles in the geometry of a rotating wormhole.

\section{Static wormholes with redshift effects\label{secr}}

If the redshift effects are present ($A'\neq 0$), wormholes with finite positive mass and $p_r$ dying out faster than any other component of the SET may exist. In this case, however, the radial gravitational tidal forces, which vanish if $A'= 0$~\cite{MT}, constrain, and may prevent, traversibility of the wormhole.

Asymptotic treatment of~\eqref{b4} reveals the following results. Wormhole solutions, with finite positive mass, that are of the form
\begin{equation}\label{r0}
\ro\sim \ro_{_\infty}r^{-3-\si} \,\text{ and }\, p_r\sim p_{r_{\infty}} r^{-3-\eta}\quad (\text{as }r\to\infty)
\end{equation}
where ($\si,\eta$) are assumed to be positive numbers, may exist if
\begin{description}
  \item[] \text{type I:} $\eta - \si > 1$ ($\Rightarrow r^{-4-\si}>r^{-3-\eta}$) yielding
   \begin{equation}\label{r1}
    4p_{t}\sim b_{_\infty}\ro_{_\infty}r^{-4-\si}\Rightarrow 4p_{t_{\infty}}=b_{_\infty}\ro_{_\infty};
   \end{equation}
  \item[] \text{type II:} $0<\eta - \si \leq 1$ yielding
   \begin{multline}\label{r2}
    4p_{t}\sim [\de_{1}^{\eta - \si} b_{_\infty}\ro_{_\infty}-2(1+\eta)p_{r_{\infty}}]r^{-3-\eta}\\
    \Rightarrow 4p_{t_{\infty}}=\de_{1}^{\eta - \si} b_{_\infty}\ro_{_\infty}-2(1+\eta)p_{r_{\infty}},
   \end{multline}
   \hspace{-8.5mm} where $\de_{1}^{\eta - \si}=1$ if $\eta - \si=1$ and 0 if $0<\eta - \si < 1$;
  \item[] \text{type III:} $\eta \leq \si$ yielding
   \begin{multline}\label{r3}
    2p_{t}\sim -(1+\eta)p_{r_{\infty}}r^{-3-\eta}\\
    \Rightarrow 2p_{t_{\infty}}= -(1+\eta)p_{r_{\infty}}.
   \end{multline}
\end{description}
In all three cases, $A$ and $b$ behave asymptotically as
\begin{equation}\label{r4}
A\sim 1-\frac{b_{_\infty}}{r},\quad b\sim b_{_\infty}-\frac{8\pi\ro_{_\infty}}{\si r^{\si}}.
\end{equation}
Notice that, in the solutions of type I,  $p_r$ vanishes asymptotically faster than the other components of the SET; these are the best solutions minimizing the use of exotic matter. In the solutions of type II, the pressures have the same asymptotic behavior and vanish faster than the energy density. To the best of our knowledge, no solutions of type I and II are available in the literature. A solution of type III, with $\eta=\si=1$ and thus $p_{t_{\infty}}=-p_{r_{\infty}}$, was derived in Eqs.~(35) to~(40) of Ref.~\cite{LPR}.

\begin{figure}
\centering
  \includegraphics[width=0.45\textwidth]{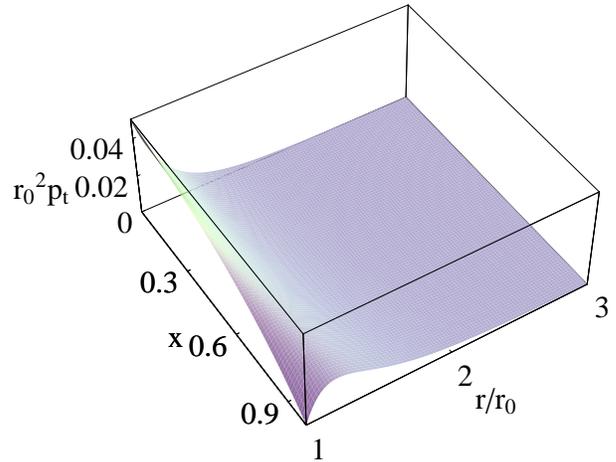}\\
  \caption{\footnotesize{A surface plot of $r_0{}^2p_t$ against ($x,r/r_0$) for the case $n=6$ [Eq.~\eqref{r20}]. Here $x\equiv 8 \pi  r_0{}^2 \rho _0=(2 M-r_0)/r_0$ and $y\equiv r/r_0$.}}\label{Fig1}
\end{figure}
\begin{figure}
\centering
  \includegraphics[width=0.45\textwidth]{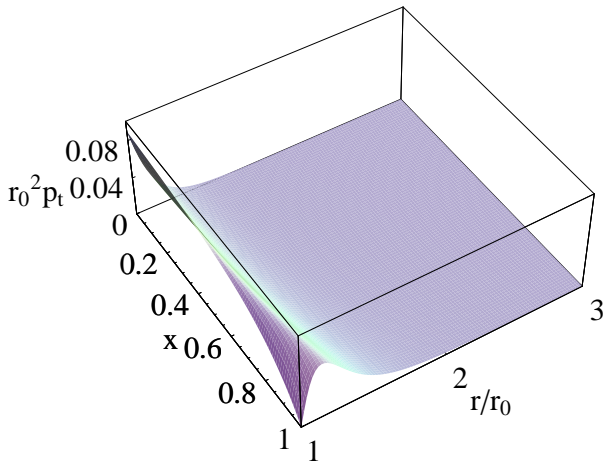}\\
  \caption{\footnotesize{A surface plot of $r_0{}^2p_t$ against ($x,r/r_0$) for the case $n=10$ [Eq.~\eqref{r21}]. Here $x\equiv 8 \pi  r_0{}^2 \rho _0=(2 M-r_0)/r_0$ and $y\equiv r/r_0$.}}\label{Fig2}
\end{figure}

Since in the remaining parts of this work we will be focusing on wormholes with positive energy density, we consider the case of wormholes with $\ro_{_\infty}>0$. Whatever the signs of the radial and transverse pressures, the classification made in Eqs.~\eqref{r1}, \eqref{r2}, and~\eqref{r3} yields the following order relations asymptotically ($r\to\infty$)
\begin{align}
&\text{type I:}& &\ro >|p_t|>|p_r|;\nn\\
\label{lec}&\text{type II:}& &\ro >|p_t|\sim |p_r|;\\
&\text{type III:}& &\ro <|p_t|\sim |p_r|.\nn
\end{align}
Using~\eqref{ec}, it is easy to see that all the LEC's are satisfied asymptotically by type I and II wormholes. If $p_{r_{\infty}}>0$ and $p_{t_{\infty}}>0$, then all the LEC's, but the DEC, are satisfied asymptotically by type III wormholes. Now, if $p_{r_{\infty}}<0$ or/and $p_{t_{\infty}}<0$, none of the LEC's are satisfied by type III wormholes.

The aim of this section is to derive closed-form wormhole solutions of type I and II. The barotropic equations~\eqref{b5} no longer are suitable as ansatzes, so we will introduce a new systematic approach.

Since the number of unknown functions in~\eqref{b4} exceeds the number of equations, we recall that the general procedure used in the literature consists in fixing two of the unknown functions and solving for the remaining functions. In the following we will fix ($\ro >0,p_r$). Notice that the classification made in Eqs.~\eqref{r1}, \eqref{r2}, and~\eqref{r3} is merely based on the asymptotic behavior of the wormholes, which is a common fact, and not on the detailed solutions or on the working ansatzes.

The aim of the following section is to use a type I wormhole, instead of a type III one~\cite{sh}, to evaluate the shadow of the SMBH candidate located at the center of the Milky Way. This is based on the criterium that type I wormholes are more realistic than type III ones having the same energy density, for the former solutions minimize the use of exotic matter. The shadow of the same SMBH has been evaluated using the Schwarzschild solution, which is the simplest known black hole solution. We will follow the same path and select the simplest type I wormhole solutions, which are derived assuming a smooth energy density distribution~\cite{MT,LPR} $\ro=\ro_0r_0{}^{m}/r^{m}$ ($m=3+\si$ and $\ro_{_\infty}=\ro_0r_0{}^{m}$). We start with the case $m=4$ ($\si =1$):
\begin{equation}\label{r4b}
    \ro=\frac{\ro_0r_0{}^4}{r^4}=\frac{\ro_{_\infty}}{r^4}\qquad (\si=1).
\end{equation}
Introducing the variable $x$ defined in~\eqref{cm1b}, the first line~\eqref{b4} yields
\begin{equation}\label{r5}
    b=(1+x)r_0-\frac{xr_0{}^2}{r}\quad\text{with}\quad 0<x\leq 1,
\end{equation}
from which we obtain
\begin{equation}\label{r5a}
   b_{_\infty}=(1+x)r_0=2M,
\end{equation}
and then
\begin{equation}\label{r6}
    b=2M-\frac{(2 M-r_0) r_0}{r}\quad\text{with}\quad x=\frac{2 M-r_0}{r_0}.
\end{equation}
Now, the constraints $0<x\leq 1$ lead to
\begin{equation}\label{r7}
    \frac{r_0}{2}<M\leq r_0
\end{equation}
where the lower limit has been shown to apply to all wormholes having finite mass and positive energy density~\eqref{cm1} and the upper limit is specific to~\eqref{r4b}. However, the choice~\eqref{r4b} is widely used in the literature~\cite{MT,LPR}. Thus, the upper limit derived in~\eqref{r7}, which results from a mere realization of the constraints~\eqref{b2} on $b$, applies to a wider set of wormhole solutions of the three types.

The next step is to choose a form for $p_r$ yielding a type I solution and determine $A$. Seeking simplicity of the final closed-form solutions, our approach consists in taking $p_r$ as a two-term polynomial in $1/r$ of the form
\begin{equation}\label{r8}
    p_r=\frac{c_n}{r^n}+\frac{c_{n+1}}{r^{n+1}}\qquad (n=3+\eta>3),
\end{equation}
yielding
\begin{multline}\label{r9}
\frac{8 \pi  r^3 p_r+b}{r (r-b)}=\frac{8 \pi(c_{n+1}+c_n r)+r_0 [(1+x) r-xr_0] r^{n-3}}{(r-r_0) (r-xr_0)
r^{n-2}}\\=\frac{N(r)}{(r-r_0) (r-xr_0)
r^{n-2}}.
\end{multline}
In order to not have a horizon at $r_0$ we set $N(r_0)\equiv 0$, where $N(r)$ is the numerator of the r.h.s. of~\eqref{r9}
\begin{equation}\label{r10}
   8 \pi(c_{n+1}+c_n r_0)+r_0{}^{n-1}=0.
\end{equation}
The remaining equation could be integrated and leads to no horizon at $r_0$ if $(0<)\,x<1$. In this case, one of the constants ($c_n,c_{n+1}$) remains undetermined and the case $x=1$ would not yield a wormhole solution.

There are two possible directions which we shall follow: Case (1), treated in Sec.~\ref{secr1}, one may add another constraint to fix both constants ($c_n,c_{n+1}$). To ease the calculations and obtain a simple closed-form metric and SET, it would be better to set $N(xr_0)= 0$, which would allow for an equal treatment of the cases $x<1$ and $x=1$ and yields a polynomial in $1/r$ in the r.h.s. of~\eqref{r9} if $n$ is an integer. For $x=1$, the constraint $N(xr_0)= 0$ is the same as $N'(r_0)= 0$ allowing $N(r)$ to have a double root at $r=r_0$, as is the denominator of the r.h.s. of~\eqref{r9}. Case (2), treated in Sec.~\ref{secr2}, one adds no further constraint. We will use $c_n$ as a free parameter and the wormhole solution will be valid only for $(0<)x<1$.

\subsection{Case (1): A further constraint ($\pmb{0<x\leq 1}$)\label{secr1}}

The constraint $N(xr_0)= 0$ reads
\begin{equation}\label{r11}
   8 \pi(c_{n+1}+c_n xr_0)+x^{n-1}r_0{}^{n-1}=0.
\end{equation}
Equations~\eqref{r10} and~\eqref{r11} are linear in ($c_n,c_{n+1}$), so one can always solve them analytically. They are identical for $x=1$. We solve them for $x<1$
\begin{equation}\label{r12}
\hspace{-1mm}c_n=-\frac{1-x^{n-1}}{8\pi (1-x)}r_0{}^{n-2},\, c_{n+1}=\frac{x(1-x^{n-2})}{8\pi (1-x)}r_0{}^{n-1},
\end{equation}
and we analytically extend them to $x=1$ since the limit, as $x\to 1$, in each r.h.s. of~\eqref{r12} exists. If $n$ is an integer\footnote{If $n$ is not an integer, the extension is still possible on introducing the two variable function $S(x,u)=(1-x^{u+1})/(1-x)$ if $x<1$ and $S(1,u)=u+1$ where $u$ is some positive reel number. This also leads to integrable expressions, but sometimes sizeable, for $A$ and the components of the SET. So, we will not consider this extension here.}, the extension is done on introducing the partial sums
\begin{equation}\label{r13}
    S_k(x)\equiv \sum_{i=0}^kx^i=\Bigg\{
                                \begin{array}{ll}
                                  \dfrac{1-x^{k+1}}{1-x}, & \text{if }\; x<1, \\
                                  k+1, & \text{if }\; x=1,
                                \end{array}
\end{equation}
($S_0(x)\equiv 1$) and the extended expressions of ($c_n,c_{n+1}$) read
\begin{equation}\label{r14}
\hspace{-1mm}c_n=-\frac{S_{n-2}(x)r_0{}^{n-2}}{8\pi},\, c_{n+1}=\frac{xS_{n-3}(x)r_0{}^{n-1}}{8\pi}.
\end{equation}
For $x=1$, these expressions coincide with those we would obtain on solving~\eqref{r10}, $N(r_0)= 0$, along with $N'(r_0)= 0$.

From now on we will omit to write the argument of $S_k$ unless there is a confusion. We can now write explicitly the expression of $p_r$
\begin{multline}\label{r15}
p_r=-\frac{S_{n-2}r_0{}^{n-2}}{8\pi r^n}+ \frac{xS_{n-3}r_0{}^{n-1}}{8\pi r^{n+1}}\\
=-\frac{[(r/r_0)-1]S_{n-2}+1}{8\pi r_0{}^{2}(r/r_0)^{n+1}}<0,
\end{multline}
where we have used $xS_{n-3}=S_{n-2}-1$. This is manifestly negative for all $r\geq r_0$. With this expression of $p_r$ the factors $r-r_0$ and $r-xr_0$ in the r.h.s. of~\eqref{r9} cancel out and the remaining expression, which is equal to $A'/A$ by the second line~\eqref{b4}, reduces to a polynomial in $1/r$ given by
\begin{equation}\label{r16}
 \frac{A'}{A}=\sum_{i=1}^{n-3}\frac{S_ir_0{}^i}{r^{i+1}}.
\end{equation}
Using the first constraint in~\eqref{b2}, we are led to
\begin{equation}\label{r17}
A=\exp\Big(-\sum_{i=1}^{n-3}\frac{S_ir_0{}^i}{i\,r^{i}}\Big).
\end{equation}
In the limit $r\to\infty$, we obtain
\begin{equation}\label{r18}
    A\sim 1-\frac{S_1r_0}{r}=1-\frac{2M}{r},
\end{equation}
where we have used~\eqref{r5a}.

Introducing the dimensionless variable $y\equiv r/r_0$ [already used in~\eqref{r15}] and re-expressing~\eqref{r4b} as $\ro=x/(8\pi r_0{}^2y^4)$, Eq.~\eqref{b4} yields
\begin{multline}\label{r18b}
  p_t=\frac{2[(n-2)y+1-n]S_{n-2}+2(n-1)}{32\pi r_0{}^2y^{n+1}}\\
+\frac{[xy^{n-3}-(y-1)S_{n-2}-1](\sum_{i=1}^{n-3}\frac{S_i}{y^{i}})}{32\pi r_0{}^2y^{n+1}}.
\end{multline}
This is a polynomial in $1/r$ the highest power of which is $xS_{n-3}{}^2/(32\pi r_0{}^2y^{2(n-1)})$ and its lowest power depends on $n$. The solution is of type III if $n=4$ ($\eta=1$), of type II if $n=5$ ($\eta=2$), and of type I if $n\geq 6$ ($\eta\geq 3$). Given $b_{_\infty}=S_1r_0$, $\ro_{_\infty}=xr_0{}^2/(8\pi)$, and $p_{r_{\infty}}=-S_{n-2}r_0{}^{n-2}/(8\pi)$, it is straightforward to check Eqs.~\eqref{r1}, \eqref{r2}, and~\eqref{r3}. For instance, for $n=5$ we find $p_{t_{\infty}}=(xS_1+6S_3)r_0{}^3/(32\pi)$, which is the coefficient of $1/r^5=1/r^{3+\eta}$ [the lowest power in~\eqref{r18}], this is conform with~\eqref{r2}. For $n\geq 6$ we find $p_{t_{\infty}}=xS_1r_0{}^3/(32\pi)$, which is the coefficient of $1/r^5=1/r^{4+\si}$ [the lowest power in~\eqref{r18}], this is conform with~\eqref{r1}.

In the limit $n\to\infty$, the graph of $p_r$ approaches that of the semi-step function $\Ta(r)$ defined by
\begin{equation}\label{stf}
    \Ta(r)=\bigg\{
                  \begin{array}{ll}
                  -1/(8\pi r_0{}^2), & \text{if }\; r=r_0, \\
                  0, & \text{if }\; r>r_0.
                  \end{array}
\end{equation}

The solution derived in this section has the property that the scaled functions ($b/r_0,r_0{}^2\ro,r_0{}^2p_r,r_0{}^2p_t$) and $A$ do depend only on ($x,y$). Using this property, it is possible to show that $r_0{}^2p_t$ may undulate for fixed $0<x\leq 1$ and $y\geq1$, as depicted in Fig.~\ref{Fig1} and Fig.~\ref{Fig2}. So, it is not possible to prove analytically the positiveness of $p_t$ because of the existence of local extreme values the critical points of which depend on $n$. On the throat, $p_t$ is positive and vanishes only in the special case $x=1$. This is obvious from its value at the point ($x,y=1$)
\begin{equation}\label{r19}
    p_t(x,1)=\frac{n-x-S_{n-2}(x)}{32\pi r_0{}^2},
\end{equation}
which is 0 if $x=1$ knowing that $S_{n-2}(1)=n-1$. For $x<1$, $p_t(x,1)> 0$ since $S_{n-2}(x)< n-1$. This is confirmed graphically for the cases $n=6$ and $n=10$, as depicted in Fig.~\ref{Fig1} and Fig.~\ref{Fig2}, where $r_0{}^2p_t$ read, respectively
\begin{align}
&r_0{}^2 p_t=\frac{x S_1}{32 \pi  y^5}+\frac{x S_2+8 S_4}{32 \pi  y^6}-\frac{9 x S_3+S_1 S_4}{32 \pi  y^7}\nn\\
\label{r20} &-\frac{S_6}{32 \pi  y^8}-\frac{(1+x^4)S_3}{32 \pi  y^9}+\frac{x S_3{}^2}{32 \pi  y^{10}},\\
&r_0{}^2 p_t=\frac{x}{32 \pi  y^5} \Big(\sum _{i=1}^5 \frac{S_i}{y^{i-1}}\Big)+\frac{16 S_8+x S_6}{32 \pi  y^{10}}\nn\\
\label{r21}&-\frac{S_1 S_8+17 x S_7}{32 \pi
 y^{11}}-\frac{1}{32 \pi  y^{12}} \Big(\sum _{i=1}^6 \frac{S_{i+9}}{y^{i-1}}\Big)+\frac{xS_7{}^2}{32 \pi  y^{18}},
\end{align}
where we have used $xS_1S_3-S_2S_4=-S_6$ and $x S_2-S_4=-(1+x^4)$ in~\eqref{r20} and $x S_i S_7-S_{i+1} S_8=-S_{i+9}$ ($i:1\to 6$) in~\eqref{r21}.

Fig.~\ref{Fig1} and Fig.~\ref{Fig2} show only a portion of the $y$ axis where $p_t\geq 0$, however, we have numerically checked that the equation $p_t(\bar{r})=0$, where $p_t$ is given by~\eqref{r20} ($n=6$) or by~\eqref{r21} ($n=10$), has no root $\bar{r}> r_0$ for $0<x\leq 1$.

\begin{figure}
\centering
  \includegraphics[width=0.327\textwidth]{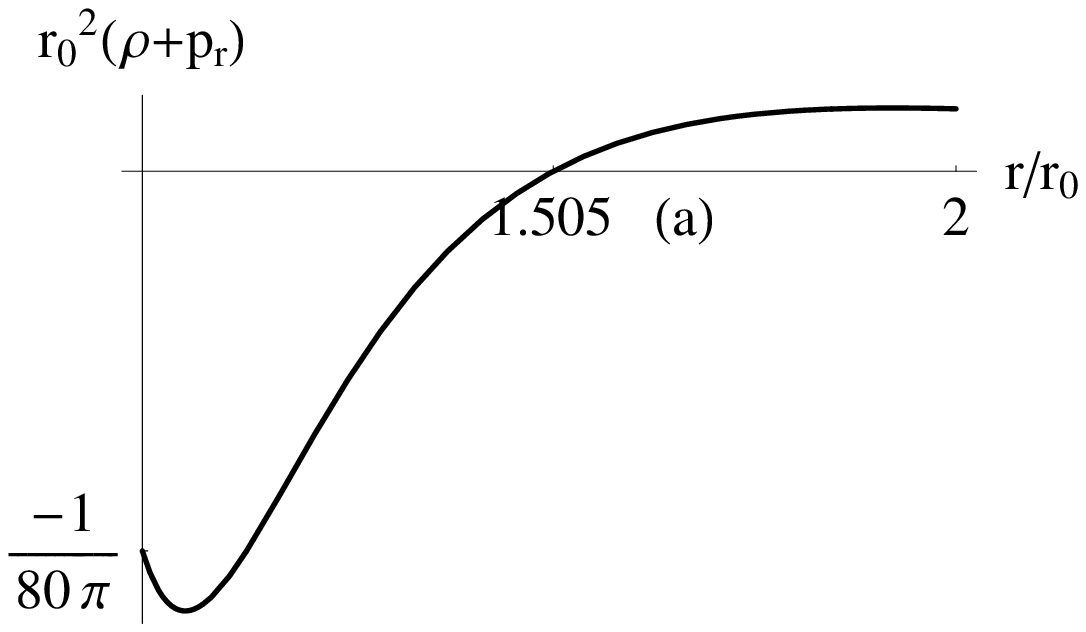} \includegraphics[width=0.327\textwidth]{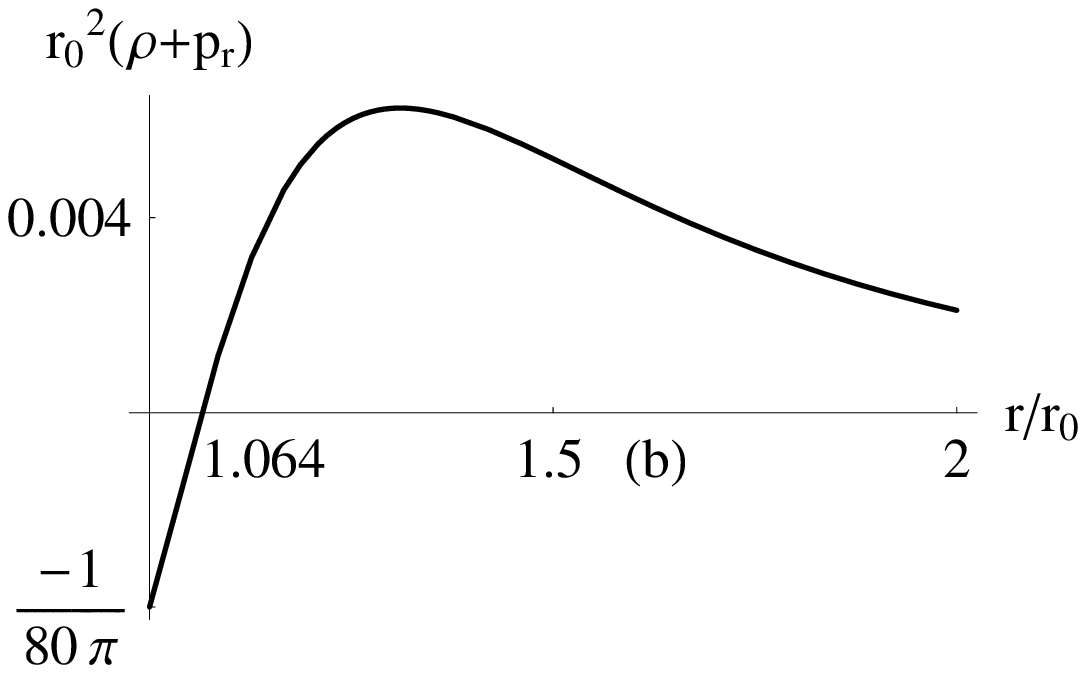}\\
  \caption{\footnotesize{Plots of $r_0{}^2(\ro +p_r)$ versus $y=r/r_0$ for (a) $n=6$ and $x=0.9$ and (b) $n=10$ and $x=0.9$. Here $\ro$ and $p_r$ are given by~\eqref{r4b} and~\eqref{r15}, respectively. We have numerically checked that the equations $\ro +p_r=0$ (for $n=6$ and $n=10$) have no roots larger than $r_0$ other than those shown in the plots. Violation of the NEC is narrowed as $n$ increases.}}\label{Fig4}
\end{figure}

Fig.~\ref{Fig4} shows clearly how the violation of the NEC is narrowed as $n$ increases. The solution we provide in the Case (2) will do better; in that, for the same value of $n$ the region of violation of the NEC gets narrower.

The proper radial distance from the throat to any point $r$, which is defined by~\eqref{prd}, takes the form
\begin{multline}
\ell =-\frac{(1+x)r_0}{2}\ln\bigg[\frac{(1-x)r_0}{(\sqrt{r-r_0}+\sqrt{r-xr_0})^2}\bigg]\\
+\sqrt{(r-r_0)(r-xr_0)}\qquad (0<x<1).
\end{multline}
If $x=1$, $\ell$ simplifies to
\begin{equation}\label{nw}
    \ell =\int_{r_0}^r\frac{\bar{r}\,\dd \bar{r}}{\bar{r}-r_0}\to\infty,
\end{equation}
which diverges. The solution corresponding to $x=1$, however it satisfies all the constraints~\eqref{b2}, it does not fulfill the requirement that it be a shortcut through spacetime between two distant asymptotically flat regions. This may not be considered as a wormhole solution.

\subsection{Case (2): No further constraints ($\pmb{0<x<1}$)\label{secr2}}

\begin{figure*}
\centering
  \includegraphics[width=0.327\textwidth]{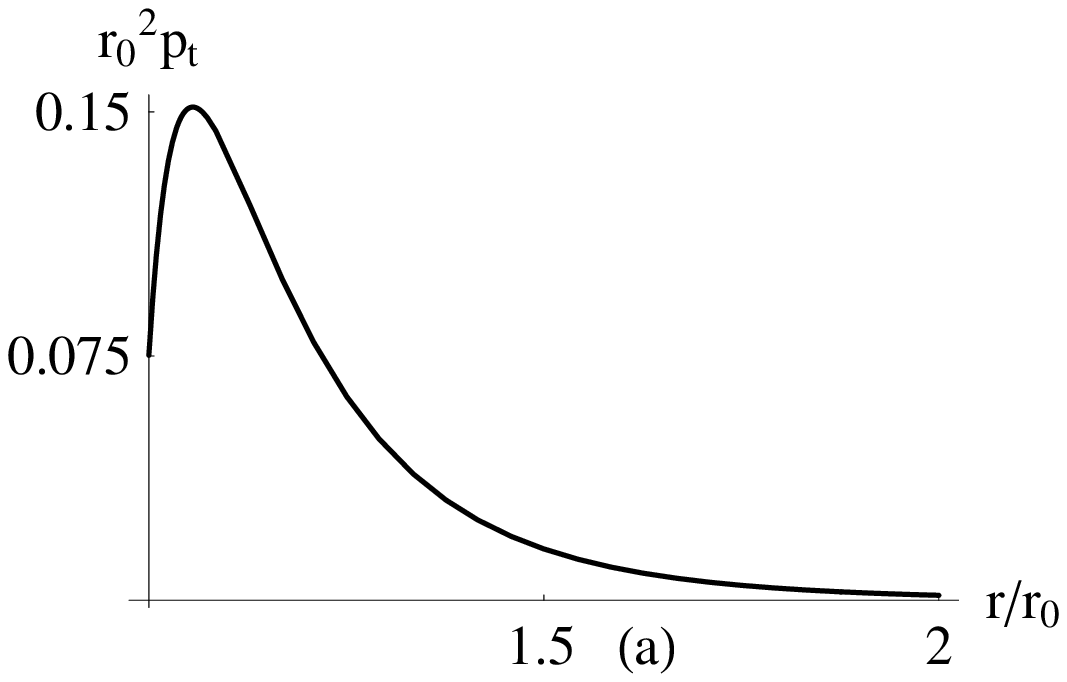} \includegraphics[width=0.327\textwidth]{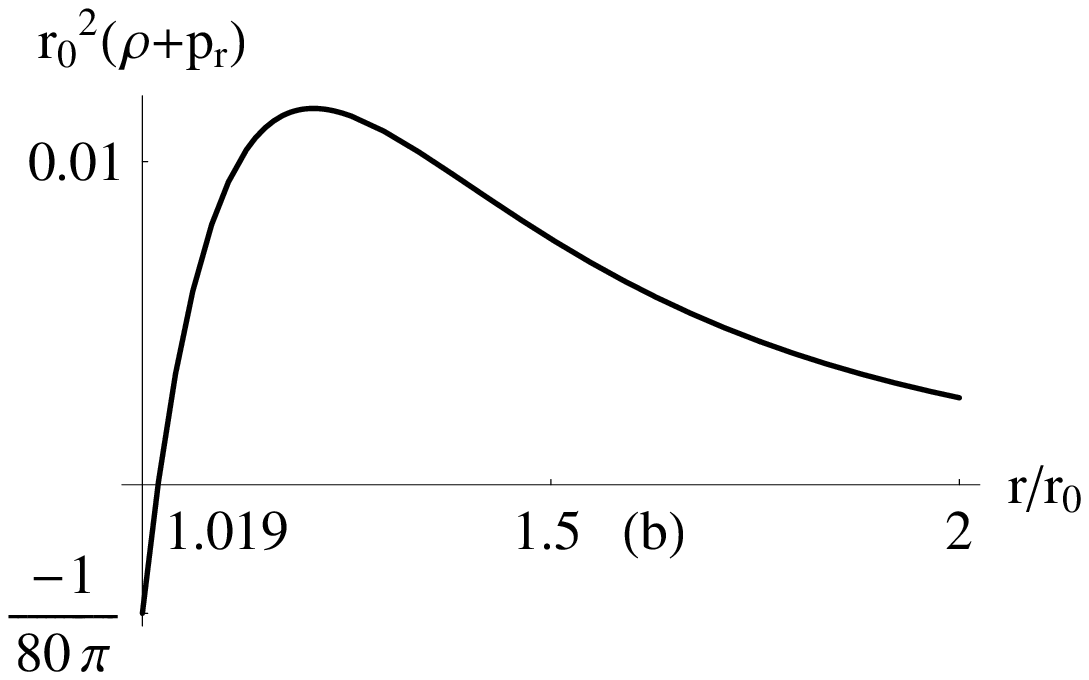} \includegraphics[width=0.327\textwidth]{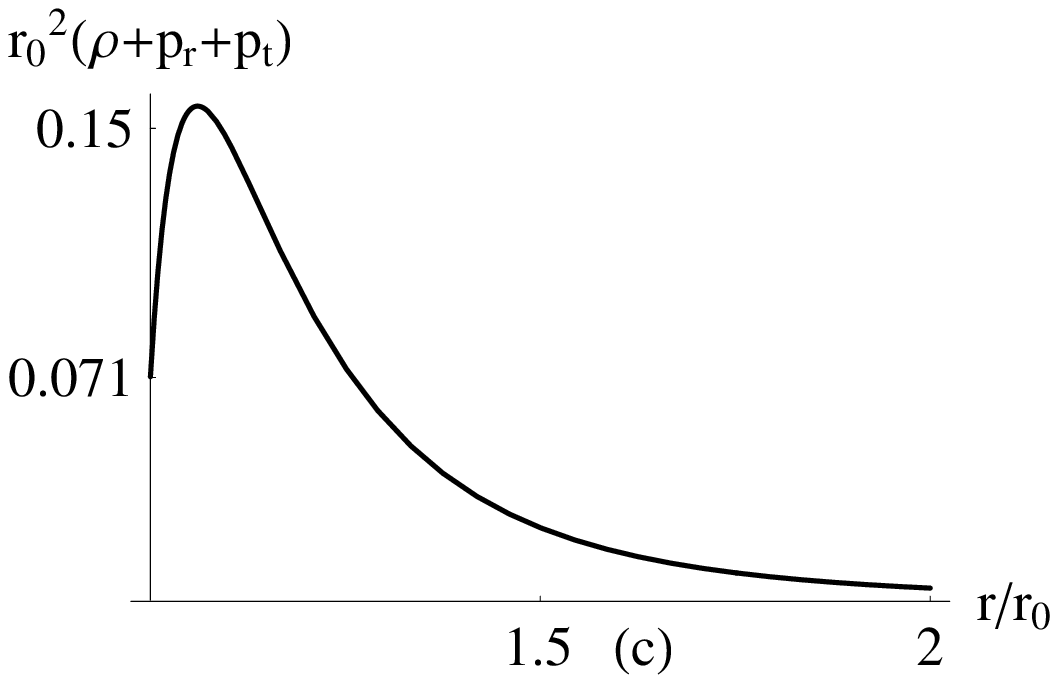}\\
  \caption{\footnotesize{Plots, for $x=0.9$ and $c=0.098$, of (a) $r_0{}^2p_t$, (b) $r_0{}^2(\ro +p_r)$, and (c) $r_0{}^2(\ro +p_r+p_t)$ versus $y=r/r_0$ where $\ro$, $p_r$, and $p_t$ are given by~\eqref{r4b}, \eqref{r2-2}, and~\eqref{r2-5}, respectively. We have numerically checked that the equations $p_t=0$ and $\ro +p_r+p_t=0$ have no roots larger than, or equal to, $r_0$ for $x=0.9$ and $c=0.098$ (see Table~\ref{Tab1}). Since $p_t$ is always positive for $x=0.9$ and $c=0.098$, the equation $\ro +p_r+2p_t=0$ also has no roots larger than, or equal to, $r_0$. The equation $\ro +p_r=0$ has a single root larger than $r_0$ given by $\bar{r}=1.019r_0$.}}\label{Fig3}
\end{figure*}

The only constraint one needs to solve is~\eqref{r10} yielding
\begin{equation}\label{r2-1}
       c_{n+1}=-r_0c_n-\frac{r_0{}^{n-1}}{8\pi}.
\end{equation}
Introducing the dimensionless constant $c$ defined by
\begin{equation}
   c_n\equiv r_0{}^{n-2}c,
\end{equation}
we obtain
\begin{equation}\label{r2-1b}
p_r= \frac{8\pi cr_0{}^{n-2}(r-r_0)-r_0{}^{n-1}}{8\pi r^{n+1}}=\frac{8\pi c(y-1)-1}{8\pi r_0{}^{2} y^{n+1}}.
\end{equation}
With this expression of $p_r$, $A$ has no horizon at $r=r_0$ for all $c_n$ at the expense of having $p_r$ negative in the vicinity of the throat: $p_r(r_0)=-1/(8\pi r_0{}^{2})$. The only constraints we may impose on $c_n$ is to ensure positiveness of $p_t$. It may seem possible to generate wormhole solutions where $p_r$ is negative only in the vicinity of the throat. In fact, from the asymptotic behavior~\eqref{r1} of type I wormholes, we see that $p_{t_{\infty}}$ is positive without constraining $c_n=p_{r_{\infty}}$. Thus, for type I wormholes, $c_n$ may a priori assume positive values, depending on $x$, provided $p_t$ remains positive everywhere. When this is the case---it is the case indeed as we shall see below---$p_r$ vanishes at some point $r_1$ then becomes positive. At $r_1$, $p_t$ is certainly positive by~\eqref{b4}. From the end-behavior~\eqref{r2} of type II wormholes, we see that $c_n$ may also assume positive values constrained by $c_n<x(1+x)r_0{}^{3}/[16\pi (n-2)]$, where we have used $b_{_\infty}=S_1r_0$, $\ro_{_\infty}=xr_0{}^2/(8\pi)$, and $n=\eta+3$.

In this Case (2), however, it is not possible to derive general solutions valid for all $n>3$~\eqref{r8}, so we provide an example of resolution for $n=6$ which will yield a type I solution:
\begin{equation}\label{r2-2}
   p_r=\frac{8\pi r_0{}^{4}c(r-r_0)-r_0{}^{5}}{8\pi r^{7}}=\frac{8\pi c(y-1)-1}{8\pi r_0{}^{2} y^{7}}.
\end{equation}
In this case also the graph of $p_r$ approaches that of the semi-step function~\eqref{stf} in the limit $n\to\infty$. With this expression of $p_r$~\eqref{r2-2} the factor $r-r_0$ in the r.h.s. of~\eqref{r9} cancels out and the remaining expression, which is equal to $A'/A$ by the second line~\eqref{b4}, reduces to
\begin{equation}\label{r2-3}
 \frac{A'}{A}=\frac{8 \pi  r_0{}^{4}c+r_0 [(1+x) r^3+r_0 r^2+r_0{}^2 r+r_0{}^3]}{r^4 (r-x r_0)}.
\end{equation}
Performing the elementary integrals, we arrive at
\begin{align}
&A=\Big(1-\frac{x r_0}{r}\Big)^{a} \exp \Big(\sum _{i=1}^3 \frac{(8 \pi  c+S_{3-i})r_0{}^i}{i\, x^{4-i}\, r^i}\Big),\nn\\
\label{r2-4}\quad &a(x,r_0,c)\equiv \dfrac{8 \pi  c+S_4}{x^4}\qquad (0<x<1),\\
&r_0^2 p_t=\frac{1}{y-x} \,\Big[\frac{x (1+x)}{32 \pi  y^4}+\frac{x-64 \pi  c}{32 \pi  y^5}\nn\\
&\qquad +\frac{10+x+8 \pi  c (11+9 x)}{32 \pi  y^6}-\frac{1+10x (1+8 \pi  c)}{32 \pi  y^7}\nn\\
&\label{r2-5}\qquad -\frac{1}{32 \pi  y^8}-\frac{1-64 \pi ^2 c^2}{32 \pi  y^9}-\frac{(1+8 \pi  c)^2}{32 \pi y^{10}}\Big].
\end{align}
The constraint $0<x<1$ yields $0<xr_0<r_0\leq r$ and this implies that $A>0$ for all $r$. The other functions ($\ro,b$) keep their expressions as given in~\eqref{r4b} and~\eqref{r5}. In the limit $r\to\infty$, ($A,p_t$) behave as in~\eqref{r18} and~\eqref{r1}, respectively.

\begin{table}[!htb]
\centering
\caption{\footnotesize Limiting values $c_{{\text{lim}}}(x)$ of $c$. The equation $p_t(\bar{r})=0$, where $p_t$ is given by~\eqref{r2-5}, has no root $\bar{r}\geq r_0$ for $c\leq c_{{\text{lim}}}(x)$. \label{Tab1}} \vspace{0.3cm}
\begin{tabular}{|||c|c||c|c||c|c|||} 
$\hspace{3mm}x\hspace{3mm}$ & $\hspace{3mm}c_{{\text{lim}}}$\hspace{3mm} & $\hspace{3mm}x\hspace{3mm}$ & \hspace{3mm}$c_{{\text{lim}}}$\hspace{3mm} & $\hspace{3mm}x\hspace{3mm}$ & $\hspace{3mm}c_{{\text{lim}}}$\hspace{3mm}  \\  \hline
 &  &  &  &  &  \vspace{-0.2cm} \\ 
0.1 & 0.012 & 0.4 & 0.037 & 0.7 & 0.069  \\ 
 &  &  &  &  &  \vspace{-0.2cm} \\ 
0.2 & 0.020 & 0.5 & 0.047 & 0.8 & 0.083  \\ 
 &  &  &  &  &  \vspace{-0.2cm} \\ 
0.3 & 0.028 & 0.6 & 0.057 & 0.9 & 0.098 \\ 
\end{tabular}
\end{table}

Table~\ref{Tab1} provides, in terms of $x$, the limiting values $c_{{\text{lim}}}(x)$ of $c$ at, or below, which $p_t$ is positive for all $r\geq r_0$. Now, for $c\leq c_{{\text{lim}}}(x)$, the transverse pressure being positive, the radial one $p_r$ is negative only near the throat, vanishes at
\begin{equation}\label{r2-6}
    r_1=r_0+\frac{r_0}{8 \pi  c},
\end{equation}
then remains positive for $r>r_1$. $r_1=1.4r_0$ for the largest value of $c_{{\text{lim}}}=0.098$ given in Table~\ref{Tab1}, corresponding to $x=0.9$, and $r_1=4.3r_0$ for the smallest one. This shows that the field equations~\eqref{b4} admit a simple, with no gluing process, wormhole solution satisfying all requirements~\eqref{b2} where the exotic matter can be made confined in a region around the throat not exceeding 1.4 times the radius of the latter.

Moreover, as Fig.~\ref{Fig3} depicts, the requirement $\ro +p_r\geq 0$ imposed by the NEC, WEC, and SEC and the requirement $p_r\in[-\ro,\ro]$ imposed by the DEC~\eqref{ec}, are violated only within a layer, adjacent to the throat, of outer and inner radii $r_{\text{out}}=1.019r_0$ and $r_0$, respectively, and all the other requirements imposed by the LEC's are satisfied by the wormhole solution corresponding to $x=0.9$ and $c=0.098$. We have thus reached the conclusion that violations of the LEC's occur partly in a narrow spherical layer [of relative extent $\ep \equiv (r_{\text{out}}-r_0)/r_0=0.019$] around the throat that might not be cumbersome for an extended object crossing the throat.

This relative extent of 0.019 can be improved, in that, reduced to much lower values on increasing $n$ and, most likely, $x$ too. In fact, for a generic value of $n$, we obtain using~\eqref{r2-1b} and $\ro=x/(8\pi r_0{}^2y^4)$
\begin{equation}\label{r2-7}
 \ro +p_r=\frac{xy^{n-3}+  8\pi cy-8\pi cy-1}{8\pi r_0{}^{2} y^{n+1}}.
\end{equation}
As is with the case $n=6$, for some values of ($c,x$) the equation $\ro +p_r=0$ has a root in the vicinity of, but larger than, 1. Taking $y=1+\ep$, we find for large $n$
\begin{equation}
 \ep \simeq \frac{1-x}{(x+8\pi c)(n-3)}.
\end{equation}
Here ($c,x$) are chosen so that the corresponding expression of $p_t$ is positive for $r\geq r_0$. We see that the size of the layer where the LEC's are violated shrinks to 0 as $n$ and $x$ increase.

The determination of solutions for $n>6$ proceeds the same way. The integrals leading to $A$ [see Eq.~\eqref{r2-3}] are all elementary of the form $\int\dd y/[y^s(y-x)]$, thus closed-form expressions for ($A,p_t$), however sizeable, are always obtainable. In fact, the general expression of $A'/A$ is brought to
\begin{equation}\label{r2-8}
  \frac{A'}{A}=\frac{8\pi c+xy^{n-3}+S_{n-3}(y)}{r_0y^{n-2}(y-x)},
\end{equation}
where $S_{n-3}(y)=\sum_{i=0}^{n-3}y^i$. Let us see the effect of large values of $n$ on traversability. The main constraint regarding traversibility is expressed in Eq.~(49) of Ref.~\cite{MT} which we re-write using our notation
\begin{multline}\label{r2-9}
 \Big|\Big(1-\frac{b}{r}\Big)\Big[-\frac{(A'/A)'}{2}+\frac{b'r-b}{2r(r-b)}\frac{(A'/A)}{2}-\frac{(A'/A)^2}{4}\Big]\Big|\\
\lesssim \frac{g_{_{\oplus}}}{(2\,\text{m})c_{\text{light}}^2}\cong \frac{1}{10^{10}\,\text{km}^2}.
\end{multline}
Here 2 m is the size of the crossing observer. Using~\eqref{r2-8} and~\eqref{r5}, the l.h.s. of~\eqref{r2-9} reduces to
\begin{equation}\label{r2-10}
   \frac{x(n-2+8\pi c+x)}{2(1-x)r_0{}^{2}},
\end{equation}
on the throat. Here $g_{_{\oplus}}$ is the value of the acceleration due to Earth gravity. For $n=6$, $c=0.098$, and $x=0.9$ the saturation in~\eqref{r2-9} results in $r_0\cong 5.7\times10^5$ km, which is a bit smaller than the radius of the Sun $R_{_{\bigodot}}=695,800$ km. Roughly speaking, if all other parameters are held constant, $n$ increases linearly with $r_0{}^{2}$ without modifying the value of the l.h.s. of~\eqref{r2-9}. To design a wormhole, say of two times the Sun's radius $r_0=2R_{_{\bigodot}}$, without violating the traversability condition~\eqref{r2-9} and with maximum confinement of the negative radial pressure around the throat we need to take $n\cong 24$. The condition~\eqref{r2-9} remains, however, satisfied for $n\lesssim 24$.

Here we have reached the same conclusion drawn in~\cite{FR}, in that, the geometry of the wormhole has very different length scales if the relative extent $\ep$ of the exotic matter (adjacent to the throat) assumes much smaller values. For the sake of example, compare the inverses of the relative rates of $b$~\eqref{r5} and $A$~\eqref{r2-8} on the throat to find
\begin{equation}
\frac{b}{b'}\sim r_0,\quad \frac{A}{A'}\sim \frac{r_0}{n-2}.
\end{equation}
This results in a discrepancy in the two scales if $n$ is large, which is the value ensuring maximum confinement. This discrepancy is obvious from Fig~\ref{Fig3} (b), and Fig~\ref{Fig4}, where the graph intersects the vertical axis at the same point $-(1-x)/(8\pi)$ independently of $n$. Hence, increasing $n$ will shift to the left the point of intersection with the $r/r_0$ axis and thus reduces the scale of variation of $\ro +p_r$.

\section{Type I wormholes for testing the nature of the SMBH candidates\label{secsd}}

We have seen that for large $n$, $\ep$ varies as $1/n$ to confine the violation of the NEC, and $r_0{}^2\sim M^2$~\eqref{r7} vary as $n$ to not harm the traversability condition. For SMWH this yields an inverse square law of $\ep$ versus $M$
\begin{equation}
  \text{SMWH: }\quad \ep\propto 1/M^2
\end{equation}
Notice that this statement does not depend on ($n,x$) and it may apply to all wormholes. For such large values of $r_0$ and $M$ the geometry of the SMWH, where $\ro\to\text{const.}$, $p_r\to 0$, and $p_t\to 0$, approaches that a SMBH, but the topology remains different. This has raised the question whether such two suppermassive objects (SMWH and SMBH) can be distinguished through astrophysical observations~\cite{sh,hs}.

An instance of such a SMBH is the one located at Sgr A$^\star$. The calculation of the photon trajectories~\cite{sh0}-\cite{sh3} yields the determination of the shape of the shadow of the emitting central object. For a static solution, this amounts to find the photon spheres which are unstable circular paths separating the absorbed paths (captured photons) and scattering ones. The apparent dividing line between black hole and sky is the apparent position of the photon sphere, which is the limiting value of the impact parameter $b_{\text{lim}}$ of the absorbed paths. It can be shown that $b_{\text{lim}}$ is related to the radius of the photon sphere $r_{\text{ps}}$ by (see, for instance\footnote{For a discussion using the Weierstrass elliptic functions see~\cite{light}.}, \cite{sh})
\begin{equation}\label{sm1}
b_{\text{lim}}=r_{\text{ps}}/\sqrt{A(r_{\text{ps}})}\,,\quad (\ln A)'=2/r_{\text{ps}}.
\end{equation}

For \Sd \aBH, $r_{\text{ps}}=3M$ yielding
\begin{equation}\label{sm2}
b_{\text{lim}}/M=3\sqrt{3}\simeq 5.196.
\end{equation}
For wormholes one usually takes $A=\exp(-2r_0/r)$ yielding $b_{\text{lim}}/r_0=\e\simeq 2.718$~\cite{sh}. However, given the nature of wormholes whose existence demands some amount of exotic matter which violates the NEC, we show that $A=\exp(-2r_0/r)$ is not the appropriate expression to work with, for it does always lead to type III wormholes. Recall that type I [respectively type III] wormholes violate the least [respectively the most] the LEC's. Substituting this expression of $A$ into the second line~\eqref{b4}, we obtain
\begin{equation}\label{sm3}
p_r=\frac{2r_0-b}{8\pi r^3}-\frac{r_0b}{4\pi r^4}.
\end{equation}
For massive wormholes, with finite mass parameter $M$, $b\sim 2M-k_1 r^{-\si}$ yielding, using the first line~\eqref{b4}, $\ro\sim k_2 r^{-3-\si}$~\eqref{r0}, where ($\si,k_1,k_2$) are positive constants. Hence, if (a) $r_0\neq M$, $|p_r|\propto r^{-3}>\ro$ as $r\to\infty$ (type III), if (b) $r_0= M$ and $0<\si\leq 1$, $|p_r|\propto r^{-3-\si}\sim \ro$ as $r\to\infty$ (type III), and if (c) $r_0= M$ and $\si > 1$, $|p_r|\propto r^{-4}> \ro$ as $r\to\infty$ (type III).

At spatial infinity (here the Earth's surface), where observations are performed in the absence of exotic matter, the wormhole solution selected to represent the SMWH, thought to inhabit the center of the Milky Way near Sgr A$^\star$, should be type I, which minimizes the use of exotic matter. In the previous section, we have developed enough tools to generate this class of massive solutions. We set $x=0.5$ and select the solution given by~\eqref{r16} and~\eqref{r17}, which we rewrite as
\begin{equation}\label{sm4}
 \frac{A'}{A}=\sum_{i=1}^{n-3}\frac{\sum_{k=0}^{i}x^k}{ry^{i}}\Rightarrow A=\exp\Big(-\sum_{i=1}^{n-3}\frac{\sum_{k=0}^{i}x^k}{i\,y^{i}}\Big),
\end{equation}
so that the second Eq.~\eqref{sm1} reads
\begin{equation}\label{sm5}
\sum_{i=1}^{n-3}\frac{\sum_{k=0}^{i}x^k}{y_{\text{ps}}{}^{i}}=2,
\end{equation}
where $y_{\text{ps}}=r_{\text{ps}}/r_0$. Solving numerically~\eqref{sm5} for $y_{\text{ps}}$ for different values of $n$, then substituting these values in the first Eq.~\eqref{sm1}, we find
\begin{align}\label{sm6}
& n=6:& & r_{\text{ps}}=1.63549 r_0,& & b_{\text{lim}}/M=4.36317,\nn\\
& n=10:& & r_{\text{ps}}=1.82946 r_0,& & b_{\text{lim}}/M=4.57738,\nn\\
& n=14:& & r_{\text{ps}}=1.84194 r_0,& & b_{\text{lim}}/M=4.58553,\\
& n\to\infty:& & r_{\text{ps}}=1.84307 r_0,& & b_{\text{lim}}/M=4.58603,\nn
\end{align}
where $b_{\text{lim}}/M=2b_{\text{lim}}/[(1+x)r_0]$ by~\eqref{r5a}. In the limit $n\to\infty$, the graph of $p_r$ approaches that of the semi-step function~\eqref{stf}.

The values of $b_{\text{lim}}/M$ given in~\eqref{sm6}, which have been derived using type I wormholes ($n\geq 6$) are much closer to the \abh value~\eqref{sm2} than the value of $b_{\text{lim}}/M=\e\simeq 2.7183$ derived with a type III wormhole. Since the ratio of the apparent diameters of the shadows is equal to the ratio of the $b_{\text{lim}}$'s, we have $\ta_{\text{S}}/\ta_{\text{W}}=5.196/4.58603=1.13301$ for the lowest ratio and $5.196/4.36317=1.19088$ for the highest one, where $\ta_{\text{S}}$ and $\ta_{\text{W}}$ are the diameters corresponding to the \Sd \abh and the wormhole, respectively. Now, $\ta_{\text{S}}=56 \,\mu\text{as}$, we obtain
\begin{equation}\label{sm7}
\ta_{\text{W}} = 47\,\mu\text{as} \text{ --- } 49 \,\mu\text{as}.
\end{equation}
Including the 14\% absolute uncertainty on $\ta_{\text{S}}$~\cite{sh}, which is $8\,\mu\text{as}$, we see that $\ta_{\text{S}}$ and $\ta_{\text{W}}$ overlap. The value $\ta_{\text{W}}$ also overlaps with the corresponding values of the Kerr solution as derived in~\cite{sh2}. We have thus reached the conclusion that the observation of the shadow is inconclusive, in that, the distinction between a (\Sd or Kerr) \abh and a wormhole, as harbored candidates at Sgr A$^\star$, is not possible within today's limits of the VLBI facilities, very recently the director team of which has reported a value of the diameter $\sim 50\,\mu\text{as}$~\cite{Sgr}.

The two bounds of $\ta_{\text{W}}$ for a type I wormhole in terms of $x$ are tabulated in Table~\ref{Tab2}.

\begin{table}[!htb]
\centering
\caption{\footnotesize The two bounds of the apparent diameter $\ta_{\text{W}}$ of the shadow of a type I wormhole in terms of $x$. As we saw earlier, the case $x=1$ may not be considered a wormhole solution~\eqref{nw}; however, when evaluating the diameter of the shadow we can use it as a limit case. \label{Tab2}} \vspace{0.3cm}
\begin{tabular}{||c|c||} 
$\hspace{7mm}x\hspace{7mm}$ & $\hspace{15mm}\ta_{\text{W}} (\mu\text{as})$ \hspace{15mm} \\  \hline
 &  \vspace{-0.2cm} \\ 
0.1 & 50 \text{ --- } 54  \\ 
 &  \vspace{-0.2cm} \\ 
0.5 & 47 \text{ --- } 49  \\ 
 &  \vspace{-0.2cm} \\ 
0.97 & 46 \text{ --- } 48 \\ 
 &  \vspace{-0.2cm} \\ 
1 & 46 \text{ --- } 48 \\ 
\end{tabular}
\end{table}

If the observed value of the diameter were much lower than $46\,\mu\text{as}$, say $30\,\mu\text{as}$, this would be an indication that the Sgr A$^\star$ might harbor a type III SMWH as well as large amounts of exotic matter. If that were the case, the difference in the diameters could be used as a measure of the amount of exotic matter harbored at Sgr A$^\star$.

The results~\eqref{sm6} and~\eqref{sm7} have been derived using type I wormholes ($n\geq 6$). Had we used a type II (respectively, type III) wormhole having the same mass and energy density~\eqref{r4b} we would have obtained for $n=5$~\eqref{r17} or type II wormhole, $r_{\text{ps}}=1.38278 r_0$, $b_{\text{lim}}/M=3.98673$, and $\ta_{\text{W}}= 43\,\mu\text{as}$ (respectively, for $n=4$~\eqref{r17} or type III wormhole, $r_{\text{ps}}=3 r_0/4$, $b_{\text{lim}}/M=\e\simeq 2.7183$, and $\ta_{\text{W}}= 29\,\mu\text{as}$). On comparing the shadows of the three types of wormholes having the same mass and energy density, we have achieved the main goal of this section consisting in showing that the Sgr A$^\star$ may harbor a type I SMWH instead of a SMBH.

The results of Table~\ref{Tab2} are specific to the class of wormhole solutions used for their derivation. The question remains open whether other classes of type I wormholes, different from those derived here, would yield similar results as those of Table~\ref{Tab2}.

Since the external geometric properties of SMWH and SMBH are similar, this leaves open the question whether a SMWH may evolve to a SMBH.

\section{Generalization\label{secg}}

There are two possible directions to generalize the method introduced in~\ref{secr}. One consists in generalizing the expression~\eqref{r8} of $p_r$ to
\begin{equation}
    p_r=\frac{c_n}{r^n}+\frac{c_{n+1}}{r^{n+1}}+\frac{c_{n+2}}{r^{n+2}}\qquad (n=\eta+3>3),
\end{equation}
which after imposing the constraint $N(r_0)\equiv 0$, generalizing~\eqref{r10}, yields a solution with two free parameters ($c_n,c_{n+1}$) to confine the exotic matter. We will not pursue this program here.

The second possibility amounts to consider higher values of $m$ (or $\si$)~\eqref{r4b} and to use the same expression~\eqref{r8} of $p_r$. We assume $\si >1$ and set $X=x/\si =8 \pi  r_0{}^2 \rho _0/\si$ so that $\ro =\ro_0r_0{}^{3+\si}/r^{3+\si}$ takes the form
\begin{equation}
 \ro=\frac{\si X}{8 \pi  r_0{}^2 y^{3+\si}}\qquad (\si >1),
\end{equation}
yielding
\begin{equation}
b=(1+X)r_0-\frac{Xr_0}{y^{\si}}.
\end{equation}
The last condition~\eqref{b2} leads to $X\leq 1/\si$ or, as before, $x\leq 1$. The second condition~\eqref{b2} reads
\begin{equation}
y^{1+\si}-(1+X)y^{\si}+X>0 \quad (y>1).
\end{equation}
This is satisfied because the polynomial on the l.h.s. vanishes at $y=1$, has a critical point $y_c=\si (1+X)/(\si +1)\leq 1$, and increases for $y>y_c$. The remaining conditions~\eqref{b2} are satisfied with $x\leq 1$. Eq.~\eqref{r7} generalizes to
\begin{equation}
    \frac{r_0}{2}<M=\frac{x+\si}{2\si}\,r_0\leq \frac{1+\si}{2\si}\,r_0.
\end{equation}
Thus, the mass of these wormholes does not exceed $r_0$, for $(1+\si)/(2\si)<1$.

If one wants to look for the lowest order type I wormhole one fixes $n=5+\si$ in~\eqref{r8}
\begin{equation}
    p_r=\frac{C}{8 \pi  r_0{}^2y^{5+\si}}-\frac{1+C}{8 \pi  r_0{}^2y^{6+\si}}\quad (3+\eta=5+\si>6),
\end{equation}
where we have already imposed the constraint eliminating the pole $y=1$ of $A'/A$ if $x<1$ [compare with ~\eqref{r10}]. If $\si$ is a positive integer, the remaining expression of $A'/A$ reads
\begin{equation}
\frac{A'}{A}=\frac{S_{\si+2}(y)+Xy^3S_{\si-1}(y)+C}{r_0y^4[y^{\si}-XS_{\si-1}(y)]},
\end{equation}
which yields a wormhole solution for $x<1$. This expression reduces to~\eqref{r2-3} if we take $\si=1$ and $C=8\pi c$.

For $\si=2$, this reads
\begin{equation}
\frac{A'}{A}=\frac{2S_{4}(y)+xy^3S_{1}(y)+2C}{r_0y^4[2y^{2}-xS_{1}(y)]}.
\end{equation}
Notice that if $x<1$, we have \[2y^{2}-xS_{1}(y)=2y^2-xy-x>0\] for all $r\geq r_0$ ($y\geq 1$): $A'/A$ has no more poles. The general expression of $A(x)$ is sizeable. For $x=8/15$, we obtain
\begin{multline}
\ln A=\frac{285 (C+1)}{16 y}-\frac{15 C}{8 y^2}+\frac{5 (C+1)}{4 y^3}\\+\frac{209+510 C}{16} \ln  y+\frac{5 (665+243 C)}{256} \ln \Big(\frac{3 y-2}{3}\Big)\\-\frac{3(2223+3125 C)}{256} \ln \Big(\frac{5 y+2}{5}\Big)\quad (x=8/15).
\end{multline}

For $x=1$, we have \[2y^{2}-xS_{1}(y)=(2y+1)(y-1).\]  We see that there is still a pole at $y=1$, which we need to impose a second constraint to eliminate it [compare with~\eqref{r11}]. This constraint reads $2S_{4}(1)+S_{1}(1)+2C=10+2+2C=0$ implying $C=-6$. Finally,
\begin{equation}
A=\Big(\frac{1+2 y}{2 y}\Big)^{57} \exp \Big(-\frac{30}{y}+\frac{6}{y^2}-\frac{10}{3 y^3}\Big)\quad (x=1).
\end{equation}

The expressions of $p_t$ for $x=8/15$ and $x=1$ are derived from~\eqref{b4}.

\section{Conclusion \label{secc}}

We have classified finite mass wormholes into three types, have introduced novel and generalizable methods for deriving, with no cutoff in the stress-energy or gluing, a class of each of the three wormhole types, and have shown the importance of type I wormholes. We have also shown the importance of type III solutions whether they are red-shift free or not.

Finite mass red-shift free wormholes are all type III and those endowed with redshift effects are three types.

Supermassive type I and type III wormholes are needed for testing whether the SMBH candidates at the center of galaxies are truly SMBH's and for computer simulations. We have shown that if the diameter of the SMBH candidate is far below the expected value, then the candidate might be a type III SMWH and that the galaxy harbor relatively large amounts of exotic matter.

The existing up-to-date VLBI facilities do not lead to differentiate the SMBH candidate at the center of the Milky Way from a possible type I SMWH, this, however, could be done in the future~\cite{Sgr}. Other signals from the galaxy, as the motion of orbiting hot spots, may lead to draw a conclusion concerning the nature of the candidate. There are existing facilities for this purpose, as the instrument Gravity~\cite{VLT} installed at the European Southern Observatory's Very Large Telescope, but due to the high similarity of the external geometries of a SMBH and a SMWH of the same mass, this duty may not perform well in the near future.




\end{document}